\def\RELEASE{1}  %
\def\ANON{0}     %
\def\SQUEEZE{0}  %

\documentclass[sigconf,10pt]{acmart}
\PassOptionsToPackage{hyphens}{url}\usepackage{hyperref}

\usepackage[noend]{algpseudocode}
\usepackage{algorithmicx}
\usepackage{booktabs}
\usepackage{caption}
\usepackage{comment}
\usepackage[inline]{enumitem}
\usepackage{graphicx}
\usepackage{listings}
\usepackage{multirow}
\usepackage{xcolor}
\usepackage{xspace}
\usepackage{libertine}
\usepackage{microtype}
\usepackage{tikz}
\usepackage{gnuplot-lua-tikz}

\usepackage{tabularx}

\usepackage{subcaption}

\microtypecontext{spacing=nonfrench}

\hypersetup{
  pdflang={en},
  pdfdisplaydoctitle}

\definecolor[named]{OurPurple}{cmyk}{0.55,1,0,0.15}
\definecolor[named]{OurDarkBlue}{cmyk}{1,0.58,0,0.21}
\hypersetup{
  colorlinks,
  linkcolor=OurPurple,
  citecolor=OurPurple,
  urlcolor=OurDarkBlue,
  filecolor=OurDarkBlue}

\if \SQUEEZE 1
\setlist[itemize]{
  leftmargin=*,
  itemsep=2pt,
  topsep=2pt}

\fi

\def\Snospace~{\S{}}

\setcopyright{none}
\renewcommand\footnotetextcopyrightpermission[1]{} %

\if \RELEASE 1
  \def\NOTES{0}
\else
  \def\NOTES{1}
\fi
\def\NOTES{0}
\if \NOTES 1
  \newcommand{\XXX}[1]{{\color{red}{XXX {#1}}}}
  \newcommand{\antoine}[1]{{\color{teal}{[\textbf{AK:} {#1}]}}}
  \newcommand{\matheus}[1]{{\color{violet}{[\textbf{MS:} {#1}]}}}
  \newcommand{\todo}[1]{{\color{blue}{TODO: {#1}}}}
\else
  \newcommand{\XXX}[1]{}
  \newcommand{\antoine}[1]{}
  \newcommand{\matheus}[1]{}
  \newcommand{\todo}[1]{}
\fi

\newcommand{\eg}{e.g.\xspace}
\newcommand{\ie}{i.e.\xspace}

\newcommand{\fakepara}[1]{\vspace{0mm}\noindent\textbf{#1}}

\if \ANON 1
  \newcommand{\sys}{ShortKut\xspace}
\else
  \newcommand{\sys}{Palette\xspace}
\fi

\begin{document}
\date{}
\title{Generating representative macrobenchmark microservice systems from distributed traces with \sys}

\if \ANON 1
  \author{Anonymous Submission \#3 (\pageref{page:last} pages)}
  \acmSubmissionID{3}
  \acmConference[ApSys'25]{ApSys}{Oct 2025}{Seoul, South Korea}
\else
  \author{Vaastav Anand}
  \affiliation{
    \institution{Max Planck Institute for Software Systems}
    \city{Saarbr\"ucken}
    \country{Germany}
  }
  \author{Matheus Stolet}
  \affiliation{
    \institution{Max Planck Institute for Software Systems}
    \city{Saarbr\"ucken}
    \country{Germany}
  }
  \author{Jonathan Mace}
  \affiliation{
    \institution{Microsoft Research}
    \city{Seattle}
    \country{USA}
  }
  \author{Antoine Kaufmann}
  \affiliation{
    \institution{Max Planck Institute for Software Systems}
    \city{Saarbr\"ucken}
    \country{Germany}
  }
\fi

\begin{abstract}
  Microservices are the dominant design for developing cloud systems
  today. Advancements for microservice need to be evaluated in representative systems, \eg with matching scale, topology, and execution patterns. 
  Unfortunately in practice, researchers and practitioners alike often do not have access to representative systems. Thus they have to resort to sub-optimal non-representative alternatives, e.g. small and oversimplified synthetic benchmark systems or simulated system models instead.

  To solve this issue, we propose the use of distributed trace datasets, available from large internet companies,
  to generate representative microservice systems.
  To do so, we introduce a novel abstraction of a \emph{system topology} which uses Graphical Causal Models (GCMs)
  to model the underlying system by incorporating the branching probabilities, execution order of outgoing
  calls to every dependency, and execution times.
  We then incorporate this topology in \sys, a system that generates
  representative flexible macrobenchmarks microservice systems from distributed traces.
\end{abstract}
 
\maketitle

\section{Introduction}

Modern cloud systems are developed as microservice systems.
They have been adopted by many large companies such as
Facebook~\cite{huye2023lifting}, Netflix~\cite{cockcroft2016microservices}, Uber~\cite{uber2015soa}, among others~\cite{cockcroft2016evolution,mazdak2017infrastructure} due to the ability of microservices
to be developed, deployed, and scaled independently.

Validating and testing new advancements for microservices often requires developers to experiment with
interventions—that is change aspects of the system with new design decisions, 
architectural choices,
algorithms, backend components, and other strategies—in order to estimate and assess their 
impact on the system.

Ideally, developers and researchers would execute these intervention experiments
on production systems to make claims about the technique's scalability and generalizability.
However, access to production systems is limited to a select few. Even if one could procure
access, the scope of experimentation is limited to minimize disruptions.

Despite the limited access to production systems,
distributed traces are often readily available and researchers can use
them to glean insight into these systems. Distributed traces capture
rich structural and temporal
information about the execution of the system, such as latency, execution patterns,
branch probabilities, and call probabilities.

\emph{We posit that distributed traces can support general purpose intervention experiments
given the rich volume of system behavior they capture.}

However, we believe that currently there are three
key challenges that prevent distributed traces from being used for intervention experiments.
First, while distributed traces capture a variety of metrics, they are observations
of the intrinsic behavior of the system. The system behavior that caused these observations
is not explicitly captured in distributed traces. Thus, we need a mechanism that
can use the distributed traces to derive a model of the intrinsic behavior of the system.
Second, currently there is no mechanism to convert intervention experiments into targeted
modifications of the system without modifying the rest of the system behavior.
The fundamental tenet of an intervention experiment requires that all the other factors
in the system must be held constant to isolate the impact of the intervention.
Third, converting distributed traces to a runnable system for supporting general purpose
intervention experiments is non-trivial as different intervention experiments might wish
to preserve different characteristics of the systems.

To overcome these challenges, we propose \sys, a system
designed for natively supporting intervention experiments using distributed traces.
\sys provides a new abstraction called \emph{system topology} that
models system behavior captured by generating a graphical causal model (GCM). 
To support targeted modifications for interventions, 
\sys provides a primitive set of operations that allow developers to make
localized changes in the system topology for a given intervention.
Finally, \sys provides a generation mechanism for converting a
system topology into runtime components.
Our proposed solution leverages the GCM at runtime to model 
the causal dependencies
from the original system and uses it to more faithfully 
sample metrics for execution behavior, such as the amount of 
work to be done in a service or the payload size to be used 
between two services given the current state of the system.

\section{Background and Motivation}%
\label{sec:bg}

\subsection{Research Use Cases}

Microservices have a large design space owing to its hetereogeneous nature. 
Consequently, the set of all possible interventions a researcher can perform is also large.
For instance, researchers building a new network stack may be interested in
confirming that their design maintains low tail latency under load so that 
requests meet their SLOs. To test their hypothesis they need a system that
matches the characteristics of a real system such as request size,
service execution times, and call graph depth and width.
Request size matters because systems optimize differently based
on its average. Service execution is critical because services with long execution times
may see negligible benefits from a faster network stack. Similarly, benefits
may be amplified in deep call graphs or offset by slow services in the critical path.
Therefore, generated systems should be able to modify and preserve these properties
so that interventions can be added while still maintaining realistic performance
characteristics.
Different use-cases, as evidenced in \autoref{tab:use-cases}, will care
about different properties when introducing interventions to the system. 
No single point solution system derived from traces can support 
all possible interventions correctly.

\begin{table}%
    \small
    \begin{tabular}{p{0.40\linewidth}p{0.55\linewidth}}
    \toprule
    \textbf{Use-case} & \textbf{Meaningful properties} \\
    \midrule
    $\mu$Service Performance\cite{zhang2024mucache,sriraman2018mutune,sriraman2019softsku} & Varying graph sizes, execution paths, topology\\
    Network Stacks\cite{kaufmann:tas,fried2024making,zhang2021demikernel} & Request sizes, service execution time, call depth, and call width\\
    Congestion Control\cite{kumar2020swift,zhou2021fasttune} & Varying topology, timeout values \\
    Res. Management\cite{somashekar2024oppertune,qiu2020firm} & Varying execution paths, large service graphs\\
    Tracing Framework\cite{las2019sifter,zhang2023benefit,fonseca2007x,toslali2021automating} & Varying execution paths, large number of services \\
    RPC Framework~\cite{kogias:r2p2,kalia:erpc} & Varying request sizes\\
    \bottomrule 
    \end{tabular}
    \caption{Research use-cases and the ideal properties from a benchmark system.}   
    \label{tab:use-cases}%
\end{table}

\subsection{The Role of Distributed Tracing}

Distributed tracing is a critical monitoring component in modern cloud systems. It provides 
troubleshooting support for developers and operators during incident root-cause analysis and post-mortems.
Distributed tracing supports this by generating an execution trace of each request across all components of the
system.
We believe that the data-rich nature of distributed trace datasets represents an opportunity for addressing the structural diversity issue
in existing open-source microservice systems for research use-cases.

\fakepara{Trace Details.} A distributed trace contains the partially-ordered list of all APIs (referred to as spans)
executed by the system to service the request. For each span, the tracing framework captures the total amount of execution time
and the specific service (or component) at which the span was executed. The tracing frameworks encodes caller-callee relationships
between APIs as parent-child relationships.

\fakepara{Distributed Tracing Workload Generation.} Distributed tracing can also be used to generate realistic workloads to test solutions at large scale because of its rich collection of execution data~\cite{sajal2024traceupscaler,du2024microservicegraphgenerator}.

\subsection{Graphical Causal Models (GCMs)}

Graphical Causal Models~\cite{elwert2013graphical} are directed acyclic graphs which encode causal relationships between the different nodes in the graph. 
Each node represents a variable, \ie some observable data, and the edges represent causality. A directed edge between two nodes
signifies that a variable influences the value of the other variable.

These models are useful for understanding the causal effects of one variable on another~\cite{budhathoki2021did} 
and have been extensively used
in root cause analysis of microservice systems~\cite{xie2024cloud,pywhyRootCause,zhao2025chase,budhathoki2022causal,janzing2019causal}.
GCMs help preserve the properties of the original system
an can generate new samples (\ie traces) based on the causal structure of the system.
As GCMs internally use inferred causal effects from the observed data, any new sample generated by the GCM
will be representative of the original data.
Due to this representativeness retention property GCMs have been used in conjunction with traces for 
performing intervention experiments in simulation~\cite{zhang2023latenseer,alomar2023causalsim}.

\subsection{Existing Approaches}

\begin{figure}%
\centering%
\begin{subfigure}[b]{0.45\linewidth}%
\centering%
\includegraphics[width=\linewidth]{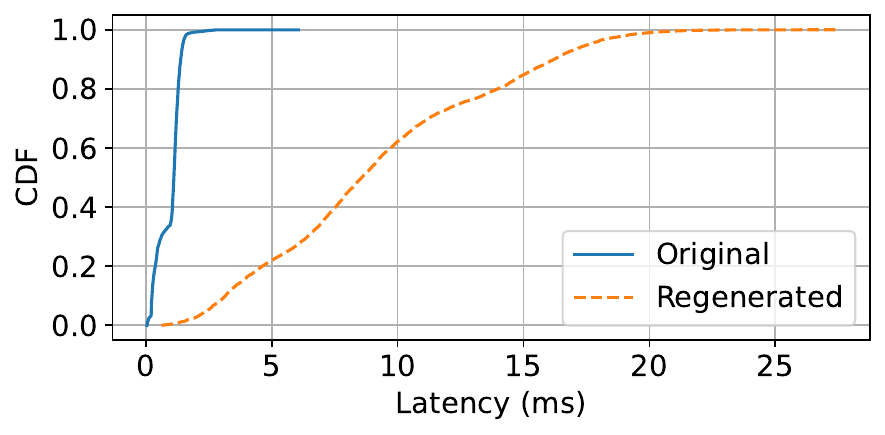}%
\caption{Service A}%
\label{fig:subfig1}%
\end{subfigure}%
\begin{subfigure}[b]{0.45\linewidth}%
\centering%
\includegraphics[width=\linewidth]{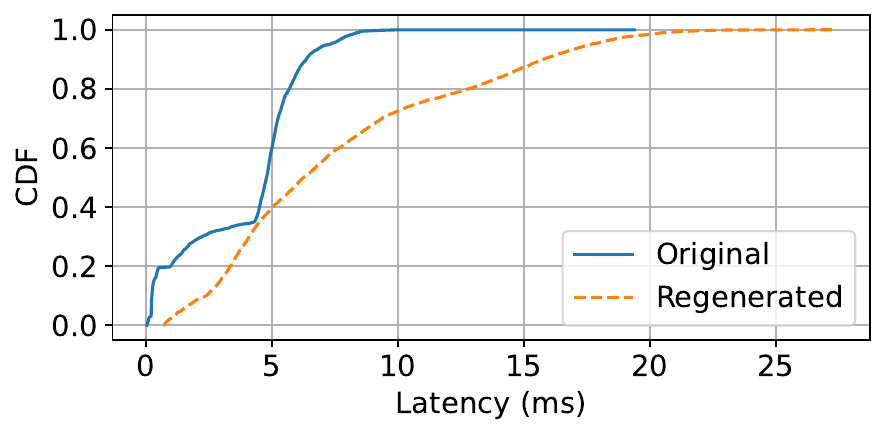}%
\caption{Service C}%
\label{fig:subfig2}%
\end{subfigure}%
\caption{Latency CDF mismatch with simple statistical approach}%
\label{fig:strawman}%
\end{figure}

\fakepara{Trace Replay.} A system that replays exact traces generated from the original system
guarantees that system behavior will be representative. However, the issue arises when a researcher performs an intervention and changes some component or aspect of the system. 
In that situation, the system will no longer have the ability to replay a prior execution because that execution is not present in the available traces.
As a result, the system could diverge from the original system beyond the true impact of the intervention.

\fakepara{Simple Statistical Models.} Another possible strawman solution is to calculate an aggregate statistic (\eg mean) for each API as well as the execution probability for each downstream dependency call. 
However, this approach fails when a downstream API has multiple callers each of which is providing a different amount of work. Let's consider the scenario with three services, where both Service A and Service C make calls to a downstream Service B to perform some amount of work.
Service C configures an order of magnitude higher work than Service A which results in Service C latency to be an order of magnitude higher than Service A.
However, when we collect statistics about the downstream Service B, this information is lost as we only get 1 statistic which is the mean.
If we were to re-generate the system using just this simple statistic, then we would find that Service B would almost take the same amount of time
for calls originating from Service C but take almost an order of magnitude higher amount of time for calls originating
from Service A. \autoref{fig:strawman} shows exactly this scenario where we find that the latency distribution of Service A is shifted by almost an order of magnitude.
The reason why this approach fails is because the latency of service B does not condition the sampling of the latency on the upstream caller.

\fakepara{Open-source microservice benchmarks}. These open-source systems are single point solutions in the large
design space of microservice systems. While these systems~\cite{gan2019open,zhou2018poster,sockshop} are
useful targets for validation and good targets for some use-cases, the systems do not cover a representative enough design space for researchers
to explore~\cite{seshagiri2022sok}. Thus, these are only useful for the specific use cases
which do not require characteristics from the design space these systems do not represent.
For example, performing an intervention experiment to test the efficacy of a new distributed
tracing system requires the system to preserve the property of a large number of services
and large fan-ins and fan-outs that are commonly seen in production systems~\cite{lee2024tale,huye2023lifting,luo2021characterizing,seemakhupt2023cloud,zhang2022crisp}.
However, this is not possible with existing open source systems as even the largest available
open-source microservice system, TrainTicket~\cite{zhou2018poster}, has only 45 services excluding caches and databases.

\section{\sys Design}%
\label{sec:design}

\begin{figure*}[t]%
\centering%
\includegraphics[width=0.95\linewidth]{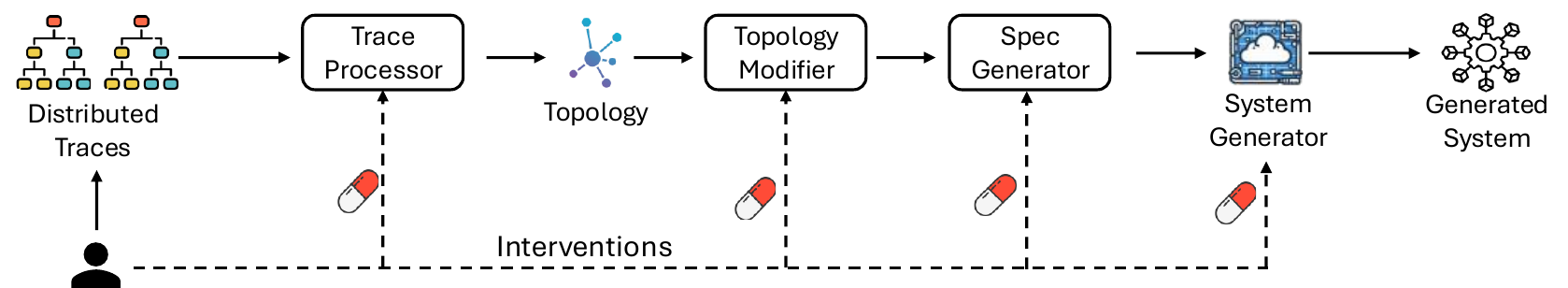}
\caption{\sys Pipeline}%
\label{fig:design}%
\end{figure*}

\autoref{fig:design} shows the design pipeline of \sys.
\sys processes trace datasets to generate a system topology that encodes structural relationships, 
execution patterns, and performance characteristics of the observed
system. This topology is converted into specifications that encodes mechanisms
to ensure the preservation of the learned characteristics.
\sys then uses Blueprint~\cite{anand2023blueprint}
to generate a full implementation of the system from these specifications.

The generated system is augmented with a GCM-based runtime
that steers execution to reflect the operation of the original
system. 
Our proposed system supports interventions by allowing users
to modify these abstractions to tailor the generated system
to their experimental needs, while still being representative
of the original system.

\subsection{System Topology}

\begin{table}[t]%
    \small
    \begin{tabularx}{\linewidth}{lX}
    \toprule
    \textbf{Behavior} & \textbf{Causal Equation}\\
    \midrule
    Probability($c$) & $p_a \sim Bernoulli(c)$ \\
    Sequential$(a_1,\dots,a_n)$ &  $p_{a_1} * \lambda_{a_1} * a_1 + \dots + p_{a_n} * \lambda_{a_n} * a_{n} + C$\\
    Concurrent$(a_1,\dots,a_n)$ & $max(p_{a_1} * \lambda_{a_1} * a_1, \dots, p_{a_n} * \lambda_{a_n} * a_{n}) + C$\\
    Choice$(a_1,\dots,a_n)$ & $p_{a_1} * \lambda_{a_1} * a_1 + \dots + p_{a_n} * \lambda_{a_n} * a_{n} + C$, such that, $p_{a_1} + \dots + p_{a_n} = 1$\\
    \bottomrule
    \end{tabularx}
\caption{Causal Equation for modeling latencies used for a given GCM node}%
\label{tab:executions}
\end{table}

\sys generates a system topology from the statistics collected during trace processing;
the topology encodes the calculated statistical information into an abstract representation
which can be exposed to the user for further modification. It models the structural
relationships within the system through a directed graph that encodes the caller-callee interactions
between services and APIs, while a Probabilistic Finite Automaton (PFA) captures
the diverse execution pathways of each API. The topology models the performance properties
of the system by using a graphical causal model (GCM). Both structural and performance properties of
the original system must be preserved to generate and execute a new representative system.

\fakepara{Directed Graph Represents Topology.} The system topology is a directed graph, $G=(P,V,E)$, 
where $P$ is the set of all 
services in the system, $V$ is the set of all APIs in the system, and $E$ is the set of all edges between all APIs.
Each partition in the graph, $p \in P$, encodes a single service uniquely identified by its name. 
Each vertex in the graph, $v \in V$,
encodes a single unique API in the system and uniquely belongs to a single partition. 
Each vertex is uniquely identified by a combination
of its name and the partition it belongs to.
Each edge in the graph, $e \in E$, encodes a caller-callee relationship between any two APIs in the system.
Edges between vertices in the same partition encode local function calls while edges between vertices in
different partitions encode remote procedure calls.

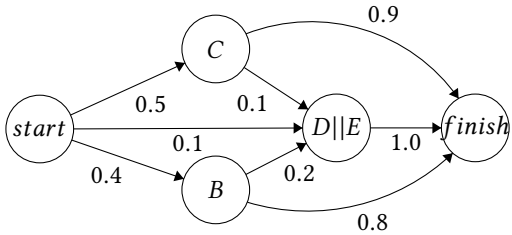
\begin{figure}
    \begin{center}
        \begin{tikzpicture}[scale=0.15]
            \tikzstyle{every node}+=[inner sep=0pt]
            \draw [black] (14.3,-21.3) circle (3);
            \draw (14.3,-21.3) node {$start$};
            \draw [black] (29.9,-26.7) circle (3);
            \draw (29.9,-26.7) node {$B$};
            \draw [black] (29.9,-14.2) circle (3);
            \draw (29.9,-14.2) node {$C$};
            \draw [black] (52.9,-21.2) circle (3);
            \draw (52.9,-21.2) node {$finish$};
            \draw [black] (40.6,-21.2) circle (3);
            \draw (40.6,-21.2) node {$D||E$};
            \draw [black] (17.03,-20.06) -- (27.17,-15.44);
            \fill [black] (27.17,-15.44) -- (26.23,-15.32) -- (26.65,-16.23);
            \draw (24.15,-18.27) node [below] {$0.5$};
            \draw [black] (17.13,-22.28) -- (27.07,-25.72);
            \fill [black] (27.07,-25.72) -- (26.47,-24.98) -- (26.15,-25.93);
            \draw (20.2,-24.57) node [below] {$0.4$};
            \draw [black] (17.3,-21.29) -- (37.6,-21.21);
            \fill [black] (37.6,-21.21) -- (36.8,-20.71) -- (36.8,-21.71);
            \draw (27.45,-21.76) node [below] {$0.1$};
            \draw [black] (32.579,-12.859) arc (111.19828:34.94669:15.952);
            \fill [black] (51.42,-18.59) -- (51.37,-17.65) -- (50.55,-18.22);
            \draw (44.76,-11.87) node [above] {$0.9$};
            \draw [black] (32.41,-15.84) -- (38.09,-19.56);
            \fill [black] (38.09,-19.56) -- (37.69,-18.7) -- (37.15,-19.54);
            \draw (33.16,-18.2) node [below] {$0.1$};
            \draw [black] (50.905,-23.436) arc (-46.35486:-106.74791:18.617);
            \fill [black] (50.9,-23.44) -- (49.98,-23.63) -- (50.67,-24.35);
            \draw (43.82,-28.73) node [below] {$0.8$};
            \draw [black] (32.57,-25.33) -- (37.93,-22.57);
            \fill [black] (37.93,-22.57) -- (36.99,-22.49) -- (37.45,-23.38);
            \draw (37.32,-24.46) node [below] {$0.2$};
            \draw [black] (43.6,-21.2) -- (49.9,-21.2);
            \fill [black] (49.9,-21.2) -- (49.1,-20.7) -- (49.1,-21.7);
            \draw (46.75,-21.7) node [below] {$1.0$};
        \end{tikzpicture}
        \end{center}
    \caption{Example PFA generated by \sys}%
    \label{fig:example_pfa}        
\end{figure}

\fakepara{PFA Encodes Execution Behavior.} 
A PFA is used to model the various execution behaviors exhibited by an API 
when invoking its dependencies. For instance, it can be used to distinguish between sequential
and concurrent calls from one API to another; this information is not captured by the directed graph
and the GCM.
\autoref{fig:example_pfa} shows an example PFA generated by \sys.
The PFA for each API consists of different states which the execution can be in. Every PFA has a start state
that represents the start of the API execution and a finish state which represents the end of the API execution.
Additionally, the PFA contains other states, each of which represents a local step in the API execution.
Each state in the PFA can make one or more concurrent outgoing calls (with some probability) to dependencies. For example, in \autoref{fig:example_pfa},
nodes B and C only make one call whereas node D$\|$E make two concurrent outgoing calls.
Once the calls end, 
the state can then transition into another state. The transition edges encode the sequential nature of
the API execution. A state may transition into one of many possible states with a different probability.
For example, in \autoref{fig:example_pfa}, the start state transitions into state B with probability 0.4, state C
with probability 0.5, and state D$\|$E with probability 0.1.
The PFA restricts state transitions such that the total probability of all outgoing transitions for a state
sum to one. The probabilistic transition between many states encodes the ability of different execution paths.

\begin{figure}[t]%
\centering%
\includegraphics[width=\linewidth]{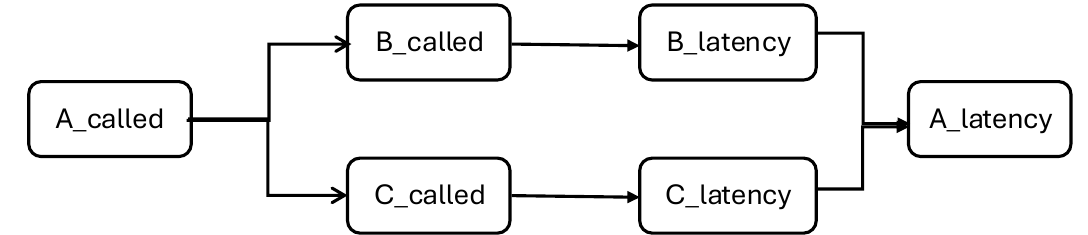}
\caption{Example latency causal graph}%
\label{fig:latency_causal_graph}%
\end{figure}

\fakepara{GCM Captures Performance Properties.}
A GCM is included in the topology structure of each API to encode the performance properties of the observed system.
Suppose we want to model the latency of an API, A, which calls two downstream APIs, B and C with differing probabilities. 
Each API's latency is represented as a separate node, and the latency of A depends on the latencies of B and C, in additional
to some local work.
\autoref{fig:latency_causal_graph} shows the causal graph for this simple scenario.
The latency of API A is directly influenced by the latencies of B and C, so there are causal edges from their latency
nodes to A's, that capture the combined impact of their latency on A's latency.
Moreover, as B and C may not always be called, this is represented as `called' nodes in the graph which represent the probability of B and C being called. Here, each `called' node can be modeled as a simple Bernoulli distribution.
However, the causal graph does not distinguish between sequential calls to B and C from concurrent calls to B and C.
This information is provided by the PFA for the API, which determines how to combine the values of the parents in the causal equation.
\autoref{tab:executions} shows the causal equation built for every latency node in the graph based on the behavior of its immediate parents.

\subsection{Generation mechanism}
\matheus{This subsection doesn't seem very interesting and I would suggest
omitting it in future versions of the paper, or at least move it to a section
other than design. Maybe implementation?}

\fakepara{Generating Topology From Traces.} \sys processes the trace dataset to calculate statistical 
information about each service from the traces
including a list of all its APIs, the execution time for each API, the probability of every outgoing call
for each API, and the dependencies for each service. The collected information can be further augmented with more
statistics depending on the information available in traces. During the processing, \sys also builds
a PFA for each API while it processes the trace data. Once the trace processing finishes, \sys
coarsens the PFA by merging similar states.
\sys then uses the built PFA to generate a causal graph for each API and generates a causal
equation for each node in the causal graph. The causal equation is entirely dependent on the
specific performance property being modeled by the GCM. \autoref{tab:executions} shows
how the equation will be built for latency nodes, the operations might be entirely different
for a different property such as payload sizes.
\sys then fits the built model to the observed trace data to find the values of the coefficients (the various $\lambda_{a}$)
to estimate the causal effect of each parent on a given node in the graph.

\fakepara{Topology to Specifications.} \sys then converts the topological model into actual source
code that encodes the performance properties of the system. For each API, \sys converts its
corresponding PFA into a set of local functions to encode PFA states and transitions. \sys also encodes a GCM model for the API and links it
to the GCM-enabled runtime. For certain performance properties, \sys also encodes how
that property should be achieved. For example, to achieve a sampled amount of latency,
\sys chooses to perform matrix multiplications for that amount of time. 

\fakepara{Implementation Generation.} We combine \sys with Blueprint~\cite{anand2023blueprint} to generate
a full implementation of the system. Blueprint provides the infrastructural pieces necessary for generating
a deployable version of the system.

\subsection{Supporting Interventions}

\sys supports interventions by allowing users to execute interventions
at four different stages.

\fakepara{Trace Processing Interventions.} Users can modify the trace
processing step to filter out invalid or irrelevant edge cases.
Moreover, users can augment the data dimensions processed by \sys
and update the GCM generation procedure to include these dimensions.
New nodes can be added
to the graph to represent new dimensions of data and they can be connected
to existing nodes that they influence. For instance, if an operator starts
collecting data on the payload size of different API calls, and the payload
size affects the latency of downstream calls, an edge between the payload
size of the calling service can be added to the vertex representing the
latency of the callee service. 

\fakepara{Topological Interventions.} The user can apply
custom modifications to the topology through a set
of modification primitives. For example, a
researcher can add new edges and vertices to represent new causal relationships
between microservices. Large topologies can also be downscaled to fit the physical
testbed available to the researcher by removing edges and vertices representing
the different APIs.

\fakepara{Specification Interventions.} The user can change how a performance property
is achieved. For example, to achieve desired latency, dependent on the use case, the user might want its services
to sleep and not do any work or they might want the services to add work that is memory bound.

\fakepara{Instantiation Interventions.} Users can add further interventions
to the system by modifying the Blueprint IR to modify concrete service
implementations. For example, the IR can be modified to support different
RPC frameworks that change how the actual benchmark system gets instantiated. 

\subsection{GCM-Based Runtime}

Each API leverages the causal equation and coefficients produced by
\sys to drive system execution in a way that resembles the
original system behavior. 
For example, the runtime samples the GCM at each API
to obtain values such as the execution time or
the payload size for outgoing messages. These samples are conditioned
on the causal relationships encoded in the GCM, allowing the system
to better capture dependencies than sampling from the observed distributions.

\fakepara{Live Measurements.} \sys makes measurements for different nodes of the causal graph during the execution and uses these
measurements into the local causal equations at every API to make predictions about the expected behavior.
For example, consider the latency
scenario from \autoref{fig:latency_causal_graph}. In the implementation of API A, \sys will insert code to
measure the latency of the outgoing calls to B and C. \sys will then plug the actual measured/observed values
into the causal equation to find the expected latency of A. 

\fakepara{Causal Baggage Propagation.} The causal graphs of some APIs might 
require measurements made higher up in the call
chain that are not directly measurable at the service executing the specific API. To ensure that all
the relevant information is correctly propagated by upstream nodes to the downstream nodes, \sys uses
baggage propagation~\cite{mace2018universal} to propagate the required measurements to the downstream
nodes executing the API. \sys can compile
and inject the exact code for adding the causal data into the baggage and propagate it
downstream along with the request because the dependency graph is known at generation time.

\fakepara{Live Corrections.} During the system execution, systems may deviate from the expected behavior of the system.
For example, a timeout may be triggered
in one of the APIs due to different performance characteristics of 
the deployed hardware. GCMs enable live corrections during runtime when execution 
diverges from the behavior observed in the original traces.
The model enables more informed sampling decisions about system
behavior by conditioning the sampling process on the underlying 
causal relationships that influence the generated metrics. 
\section{Conclusions}

In this paper, we have presented the design of an extensible system for generating
representative macrobenchmark microservice systems from distributed trace datasets
that uses Graphical Causal Models (GCMs) for modeling system behavior to
preserve the desired representative characteristics of a microservice systems.
We believe that \sys provides a flexible way for users to conduct intervention
experiments to validate and test new advancements for microservices. \if \ANON 0
\fi

\bibliographystyle{plain}
\bibliography{paper,bibdb/papers,bibdb/strings,bibdb/defs}

\begin{thebibliography}{10}

\bibitem{alomar2023causalsim}
Abdullah Alomar, Pouya Hamadanian, Arash Nasr-Esfahany, Anish Agarwal, Mohammad
  Alizadeh, and Devavrat Shah.
\newblock $\{$CausalSim$\}$: A causal framework for unbiased
  $\{$Trace-Driven$\}$ simulation.
\newblock In {\em 20th USENIX Symposium on Networked Systems Design and
  Implementation (NSDI 23)}, pages 1115--1147, 2023.

\bibitem{anand2023blueprint}
Vaastav Anand, Deepak Garg, Antoine Kaufmann, and Jonathan Mace.
\newblock Blueprint: A toolchain for highly-reconfigurable microservice
  applications.
\newblock In {\em Proceedings of the 29th Symposium on Operating Systems
  Principles}, pages 482--497, 2023.

\bibitem{budhathoki2021did}
Kailash Budhathoki, Dominik Janzing, Patrick Bloebaum, and Hoiyi Ng.
\newblock Why did the distribution change?
\newblock In {\em International Conference on Artificial Intelligence and
  Statistics}, pages 1666--1674. PMLR, 2021.

\bibitem{budhathoki2022causal}
Kailash Budhathoki, Lenon Minorics, Patrick Bl{\"o}baum, and Dominik Janzing.
\newblock Causal structure-based root cause analysis of outliers.
\newblock In {\em International conference on machine learning}, pages
  2357--2369. PMLR, 2022.

\bibitem{cockcroft2016evolution}
Adrian Cockcroft.
\newblock The evolution of microservices.
\newblock (April 2016). Retrieved October 2020 from
  \url{https://www.slideshare.net/adriancockcroft/evolution-of-microservices-craft-conference},
  2016.

\bibitem{cockcroft2016microservices}
Adrian Cockcroft.
\newblock Microservices workshop: Why, what, and how to get there.
\newblock (April 2016). Retrieved October 2020 from
  \url{https://www.slideshare.net/adriancockcroft/microservices-workshop-craft-conference},
  2016.

\bibitem{pywhyRootCause}
DoWhy documentation v0.8.
\newblock Root cause analysis (rca) of latencies in a microservice
  architecture.
\newblock
  \url{https://www.pywhy.org/dowhy/v0.8/example_notebooks/rca_microservice_architecture.html},
  2024.
\newblock [Accessed 05-06-2025].

\bibitem{du2024microservicegraphgenerator}
Fanrong Du, Jiuchen Shi, Quan Chen, Li~Li, and Minyi Guo.
\newblock A microservice graph generator with production characteristics, 2024.

\bibitem{elwert2013graphical}
Felix Elwert.
\newblock Graphical causal models.
\newblock In {\em Handbook of causal analysis for social research}, pages
  245--273. Springer, 2013.

\bibitem{fonseca2007x}
Rodrigo Fonseca, George Porter, Randy~H Katz, and Scott Shenker.
\newblock $\{$X-Trace$\}$: A pervasive network tracing framework.
\newblock In {\em 4th USENIX Symposium on Networked Systems Design \&
  Implementation (NSDI 07)}, 2007.

\bibitem{fried2024making}
Joshua Fried, Gohar~Irfan Chaudhry, Enrique Saurez, Esha Choukse, {\'I}{\~n}igo
  Goiri, Sameh Elnikety, Rodrigo Fonseca, and Adam Belay.
\newblock Making kernel bypass practical for the cloud with junction.
\newblock In {\em 21st USENIX Symposium on Networked Systems Design and
  Implementation (NSDI 24)}, pages 55--73, 2024.

\bibitem{gan2019open}
Yu~Gan, Yanqi Zhang, Dailun Cheng, Ankitha Shetty, Priyal Rathi, Nayan Katarki,
  Ariana Bruno, Justin Hu, Brian Ritchken, Brendon Jackson, et~al.
\newblock An open-source benchmark suite for microservices and their
  hardware-software implications for cloud \& edge systems.
\newblock In {\em Proceedings of the Twenty-Fourth International Conference on
  Architectural Support for Programming Languages and Operating Systems}, pages
  3--18, 2019.

\bibitem{uber2015soa}
Einas Haddad.
\newblock Service-oriented architecture: Scaling the uber engineering codebase
  as we grow.
\newblock (September 2015). Retrieved October 2020 from
  \url{https://eng.uber.com/service-oriented-architecture/}, 2015.

\bibitem{mazdak2017infrastructure}
Mazdak Hashemi.
\newblock (January 2017). Retrieved February 2021 from
  \url{https://blog.twitter.com/engineering/en_us/topics/infrastructure/2017/the-infrastructure-behind-twitter-scale.html},
  2017.

\bibitem{huye2023lifting}
Darby Huye, Yuri Shkuro, and Raja~R Sambasivan.
\newblock Lifting the veil on $\{$Meta’s$\}$ microservice architecture:
  Analyses of topology and request workflows.
\newblock In {\em 2023 USENIX Annual Technical Conference (USENIX ATC 23)},
  pages 419--432, 2023.

\bibitem{janzing2019causal}
Dominik Janzing, Kailash Budhathoki, Lenon Minorics, and Patrick Bl{\"o}baum.
\newblock Causal structure based root cause analysis of outliers.
\newblock {\em arXiv preprint arXiv:1912.02724}, 2019.

\bibitem{kalia:erpc}
Anuj Kalia, Michael Kaminsky, and David Andersen.
\newblock Datacenter {RPCs} can be general and fast.
\newblock In {\em 16th USENIX Symposium on Networked Systems Design and
  Implementation}, NSDI, 2019.

\bibitem{kaufmann:tas}
Antoine Kaufmann, Tim Stamler, Simon Peter, Naveen~Kr. Sharma, Arvind
  Krishnamurthy, and Thomas Anderson.
\newblock {TAS}: {TCP} acceleration as an {OS} service.
\newblock In {\em 14th ACM European Conference on Computer Systems}, EuroSys,
  2019.

\bibitem{kogias:r2p2}
Marios Kogias, George Prekas, Adrien Ghosn, Jonas Fietz, and Edouard Bugnion.
\newblock {R2P2}: Making {RPCs} first-class datacenter citizens.
\newblock In {\em 2019 USENIX Annual Technical Conference}, ATC, 2019.

\bibitem{kumar2020swift}
Gautam Kumar, Nandita Dukkipati, Keon Jang, Hassan~MG Wassel, Xian Wu, Behnam
  Montazeri, Yaogong Wang, Kevin Springborn, Christopher Alfeld, Michael Ryan,
  et~al.
\newblock Swift: Delay is simple and effective for congestion control in the
  datacenter.
\newblock In {\em Proceedings of the Annual conference of the ACM Special
  Interest Group on Data Communication on the applications, technologies,
  architectures, and protocols for computer communication}, pages 514--528,
  2020.

\bibitem{las2019sifter}
Pedro Las-Casas, Giorgi Papakerashvili, Vaastav Anand, and Jonathan Mace.
\newblock Sifter: Scalable sampling for distributed traces, without feature
  engineering.
\newblock In {\em Proceedings of the ACM Symposium on Cloud Computing}, pages
  312--324, 2019.

\bibitem{lee2024tale}
I-Ting~Angelina Lee, Zhizhou Zhang, Abhishek Parwal, and Milind Chabbi.
\newblock The tale of errors in microservices.
\newblock {\em Proceedings of the ACM on Measurement and Analysis of Computing
  Systems}, 8(3):1--36, 2024.

\bibitem{luo2021characterizing}
Shutian Luo, Huanle Xu, Chengzhi Lu, Kejiang Ye, Guoyao Xu, Liping Zhang,
  Yu~Ding, Jian He, and Chengzhong Xu.
\newblock Characterizing microservice dependency and performance: Alibaba trace
  analysis.
\newblock In {\em Proceedings of the ACM Symposium on Cloud Computing}, pages
  412--426, 2021.

\bibitem{mace2018universal}
Jonathan Mace and Rodrigo Fonseca.
\newblock Universal context propagation for distributed system instrumentation.
\newblock In {\em Proceedings of the thirteenth EuroSys conference}, pages
  1--18, 2018.

\bibitem{sockshop}
microservices demo.
\newblock Sockshop.
\newblock Retrieved August 2022 from
  \url{https://github.com/microservices-demo/microservices-demo}, 2016.

\bibitem{qiu2020firm}
Haoran Qiu, Subho~S Banerjee, Saurabh Jha, Zbigniew~T Kalbarczyk, and
  Ravishankar~K Iyer.
\newblock $\{$FIRM$\}$: An intelligent fine-grained resource management
  framework for $\{$SLO-Oriented$\}$ microservices.
\newblock In {\em 14th USENIX symposium on operating systems design and
  implementation (OSDI 20)}, pages 805--825, 2020.

\bibitem{sajal2024traceupscaler}
Sultan~Mahmud Sajal, Timothy Zhu, Bhuvan Urgaonkar, and Siddhartha Sen.
\newblock Traceupscaler: Upscaling traces to evaluate systems at high load.
\newblock In {\em Proceedings of the Nineteenth European Conference on Computer
  Systems}, pages 942--961, 2024.

\bibitem{seemakhupt2023cloud}
Korakit Seemakhupt, Brent~E Stephens, Samira Khan, Sihang Liu, Hassan Wassel,
  Soheil~Hassas Yeganeh, Alex~C Snoeren, Arvind Krishnamurthy, David~E Culler,
  and Henry~M Levy.
\newblock A cloud-scale characterization of remote procedure calls.
\newblock In {\em Proceedings of the 29th Symposium on Operating Systems
  Principles}, pages 498--514, 2023.

\bibitem{seshagiri2022sok}
Vishwanath Seshagiri, Darby Huye, Lan Liu, Avani Wildani, and Raja~R
  Sambasivan.
\newblock [sok] identifying mismatches between microservice testbeds and
  industrial perceptions of microservices.
\newblock {\em Journal of Systems Research}, 2(1), 2022.

\bibitem{somashekar2024oppertune}
Gagan Somashekar, Karan Tandon, Anush Kini, Chieh-Chun Chang, Petr Husak,
  Ranjita Bhagwan, Mayukh Das, Anshul Gandhi, and Nagarajan Natarajan.
\newblock $\{$OPPerTune$\}$:$\{$Post-Deployment$\}$ configuration tuning of
  services made easy.
\newblock In {\em 21st USENIX Symposium on Networked Systems Design and
  Implementation (NSDI 24)}, pages 1101--1120, 2024.

\bibitem{sriraman2019softsku}
Akshitha Sriraman, Abhishek Dhanotia, and Thomas~F Wenisch.
\newblock Softsku: Optimizing server architectures for microservice diversity@
  scale.
\newblock In {\em Proceedings of the 46th International Symposium on Computer
  Architecture}, pages 513--526, 2019.

\bibitem{sriraman2018mutune}
Akshitha Sriraman and Thomas~F Wenisch.
\newblock $\{$$\mu$Tune$\}$:$\{$Auto-Tuned$\}$ threading for $\{$OLDI$\}$
  microservices.
\newblock In {\em 13th USENIX Symposium on Operating Systems Design and
  Implementation (OSDI 18)}, pages 177--194, 2018.

\bibitem{toslali2021automating}
Mert Toslali, Emre Ates, Alex Ellis, Zhaoqi Zhang, Darby Huye, Lan Liu,
  Samantha Puterman, Ayse~K Coskun, and Raja~R Sambasivan.
\newblock Automating instrumentation choices for performance problems in
  distributed applications with vaif.
\newblock In {\em Proceedings of the ACM Symposium on Cloud Computing}, pages
  61--75, 2021.

\bibitem{xie2024cloud}
Zhiqiang Xie, Yujia Zheng, Lizi Ottens, Kun Zhang, Christos Kozyrakis, and
  Jonathan Mace.
\newblock Cloud atlas: Efficient fault localization for cloud systems using
  language models and causal insight.
\newblock {\em arXiv preprint arXiv:2407.08694}, 2024.

\bibitem{zhang2024mucache}
Haoran Zhang, Konstantinos Kallas, Spyros Pavlatos, Rajeev Alur, Sebastian
  Angel, and Vincent Liu.
\newblock $\{$MuCache$\}$: A general framework for caching in microservice
  graphs.
\newblock In {\em 21st USENIX Symposium on Networked Systems Design and
  Implementation (NSDI 24)}, pages 221--238, 2024.

\bibitem{zhang2021demikernel}
Irene Zhang, Amanda Raybuck, Pratyush Patel, Kirk Olynyk, Jacob Nelson, Omar
  S~Navarro Leija, Ashlie Martinez, Jing Liu, Anna~Kornfeld Simpson, Sujay
  Jayakar, et~al.
\newblock The demikernel datapath os architecture for microsecond-scale
  datacenter systems.
\newblock In {\em Proceedings of the ACM SIGOPS 28th Symposium on Operating
  Systems Principles}, pages 195--211, 2021.

\bibitem{zhang2023benefit}
Lei Zhang, Zhiqiang Xie, Vaastav Anand, Ymir Vigfusson, and Jonathan Mace.
\newblock The benefit of hindsight: Tracing $\{$Edge-Cases$\}$ in distributed
  systems.
\newblock In {\em 20th USENIX Symposium on Networked Systems Design and
  Implementation (NSDI 23)}, pages 321--339, 2023.

\bibitem{zhang2023latenseer}
Yazhuo Zhang, Rebecca Isaacs, Yao Yue, Juncheng Yang, Lei Zhang, and Ymir
  Vigfusson.
\newblock Latenseer: Causal modeling of end-to-end latency distributions by
  harnessing distributed tracing.
\newblock In {\em Proceedings of the 2023 ACM Symposium on Cloud Computing},
  pages 502--519, 2023.

\bibitem{zhang2022crisp}
Zhizhou Zhang, Murali~Krishna Ramanathan, Prithvi Raj, Abhishek Parwal, Timothy
  Sherwood, and Milind Chabbi.
\newblock $\{$CRISP$\}$: Critical path analysis of $\{$Large-Scale$\}$
  microservice architectures.
\newblock In {\em 2022 USENIX Annual Technical Conference (USENIX ATC 22)},
  pages 655--672, 2022.

\bibitem{zhao2025chase}
Ziming Zhao, Zhenwei Wang, Tiehua Zhang, Zhishu Shen, Hai Dong, Zhen Lei,
  Xingjun Ma, Gaowei Xu, Zhijun Ding, and Yun Yang.
\newblock Chase: A causal hypergraph based framework for root cause analysis in
  multimodal microservice systems, 2025.

\bibitem{zhou2021fasttune}
Renjie Zhou, Dezun Dong, Shan Huang, and Yang Bai.
\newblock Fasttune: Timely and precise congestion control in data center
  network.
\newblock In {\em 2021 IEEE Intl Conf on Parallel \& Distributed Processing
  with Applications, Big Data \& Cloud Computing, Sustainable Computing \&
  Communications, Social Computing \& Networking
  (ISPA/BDCloud/SocialCom/SustainCom)}, pages 238--245. IEEE, 2021.

\bibitem{zhou2018poster}
Xiang Zhou, Xin Peng, Tao Xie, Jun Sun, Chenjie Xu, Chao Ji, and Wenyun Zhao.
\newblock Poster: Benchmarking microservice systems for software engineering
  research.
\newblock In {\em 2018 IEEE/ACM 40th International Conference on Software
  Engineering: Companion (ICSE-Companion)}, pages 323--324. IEEE, 2018.

\end{thebibliography}

\label{page:last}
\end{document}